\documentclass[%
 preprint,
 amsmath,amssymb,
 aps,
 prl,%longbibliography,%onecolumn
]{revtex4-1}

\usepackage{graphicx}
\usepackage{dcolumn}
\usepackage{bm}
\usepackage{color}
\usepackage{booktabs} 
\usepackage{multirow}

\begin{document}

\title{Quantum CZ Gate based on Single Gradient Metasurface}
\author{Qi Liu$^{1,2}$}
\author{Yu Tian$^{1,2}$}
\author{Zhaohua Tian$^{1}$}
\author{Guixin Li$^3$}
\author{Xi-Feng Ren$^{4,5}$}
\author{Qihuang Gong$^{1,2,5,6,7}$}
\author{Ying Gu$^{1,2,5,6,7}$}
 \email{ygu@pku.edu.cn}

\affiliation{$^1$State Key Laboratory for Mesoscopic Physics, Department of Physics, Peking University, Beijing 100871, China\\
$^2$Frontiers Science Center for Nano-optoelectronics $\&$  Collaborative Innovation Center of Quantum Matter $\&$ Beijing Academy of Quantum Information Sciences, Peking University, Beijing 100871, China\\
$^3$Department of Materials Science and Engineering, Southern University of Science and Technology, Shenzhen 518055, China\\
$^4$CAS Key Laboratory of Quantum Information, University of Science and Technology of China, Hefei 230026, China\\
$^5$Hefei National Laboratory, Hefei 230088, China\\
$^6$Collaborative Innovation Center of Extreme Optics, Shanxi University, Taiyuan, Shanxi 030006, China\\
$^7$Peking University Yangtze Delta Institute of Optoelectronics, Nantong 226010, China
}

\date{\today}
\begin{abstract}
We propose a scheme to realize quantum controlled-Z (CZ) gates through single gradient metasurface.
Using its unique parallel beam-splitting feature, i.e., a series of connected beam splitters with the same splitting ratio, one metasurface can support a CZ gate, several independent CZ gates, or a cascaded CZ gates.
Taking advantage of the input polarization determined output path-locking feature, both polarization-encoded and path-encoded CZ gates can be demonstrated on the same metasurface, which further improves the integration level of quantum devices.
Our research paves the way for integrating quantum logical function through the metasurface. 
%????????????????????????????????CZ????????????CZ??
%it is highly possible to construct integrable multiqubit quantum gates by extending our scheme into a multilayer metasurface. Our results open an avenue for high-density and multifunction quantum integration on metasurface.

\begin{description}
\item[Keywords]  Quantum CZ gate; Metasurface;  Integrated quantum photonics
%\item[PACS numbers]
\end{description}

\end{abstract}

%\pacs{Valid PACS appear here}

\maketitle

%%%% Section I
%\textit{Introduction.}

Metasurface, emerging as a novel planar platform for light manipulation~\cite{MSR1,MSR2,MSR3,MSR4,MSR5}, offers a new paradigm for quantum integration by enabling modulation among multiple degrees of freedom~\cite{MQR1,MQR2}. 
Through the metasurface, various quantum functionalities are realized, 
including generation~\cite{MQP1,MQP2,MQP3} and manipulation~\cite{MQM1,MQM2,MQM3,MQM4} of high-dimensional entanglement, quantum tomography~\cite{MQT}, quantum imaging~\cite{MQI1,MQI2,MQI3},  quantum sensing~\cite{MQS}, and beam-splitting functions~\cite{MSBS1,MSBS2,MSBS3,MSBS4,MSBS5}. 
Recently, we have revealed the unique parallel beam-splitting feature of gradient metasurface, which further is used to prepare entangled state and realize entanglement fusion ~\cite{MPBS}, suggesting the potential for scalable quantum integration. 
Despite these progresses, the demonstration of fundamental logic operations (such as Controlled-NOT (CNOT) or Controlled-Z (CZ) gate) through metasurfaces remains an open challenge.

CZ gates, combined with single-qubit operations, can construct any quantum circuit, enabling universal quantum computation and Bell-state discrimination~\cite{CNOT_QIP1,CNOT_QIP2,CNOT_QIP3}.
Traditionally, using the principles of linear optics \cite{KLM,SKLM1,SKLM2,SKLM3}, quantum CZ gates can be experimentally validated with several identical beam splitters~\cite{SKLM4,SKLM5,SKLM6,SKLM7,SKLM8}. 
However, the scalability of quantum logic gates is constrained by the size of those beam splitters. 
Since 2008, various on-chip CZ gates have been realized for both path-encoded~\cite{CHIP1,CHIP2,CHIP3} and polarization-encoded~\cite{CHIP4,CHIP5} qubits, with gate dimensions reduced to hundreds of micrometers. 
With plasmonic structures~\cite{COMPACT1} and novel waveguide coupling techniques~\cite{COMPACT2}, 
the size of CZ gate can be further reduced.
These innovations, along with symmetry-breaking~\cite{COMPACT3} and optical inverse design~\cite{COMPACT4} methods, have miniaturized CZ gates to near-wavelength dimension. 
%?????CZ??????????????????????????????????????????????????CZ???????????????????????????????????????????????????????????????????????????????????CZ?????????????????????????????

One quantum CZ gate needs at least three identical beam-splitters with the same splitting ratio.
While a quantum logical device generally requires a set of CZ gates, i.e., many beam splitters,
if they are fabricated by micro/nano structures, 
the factors such as asymmetric, crosstalk, and loss of beam splitters, coupled with post-selection processes, bring great uncertainty to the quantum logic function.
So using one micro/nano photonic structure to realize the CZ gate is very attractive, but it has never been proposed.
According to the recent finding that gradient metasurface has the peculiarity of parallel beam-splitting, i.e., a set of connected beam-splitters with the same splitting ratio~\cite{MPBS},  we propose a scheme to realize quantum CZ gates through single gradient metasurface. 

\begin{figure}[tbp]
  \centering
  \includegraphics[width=0.6\textwidth]{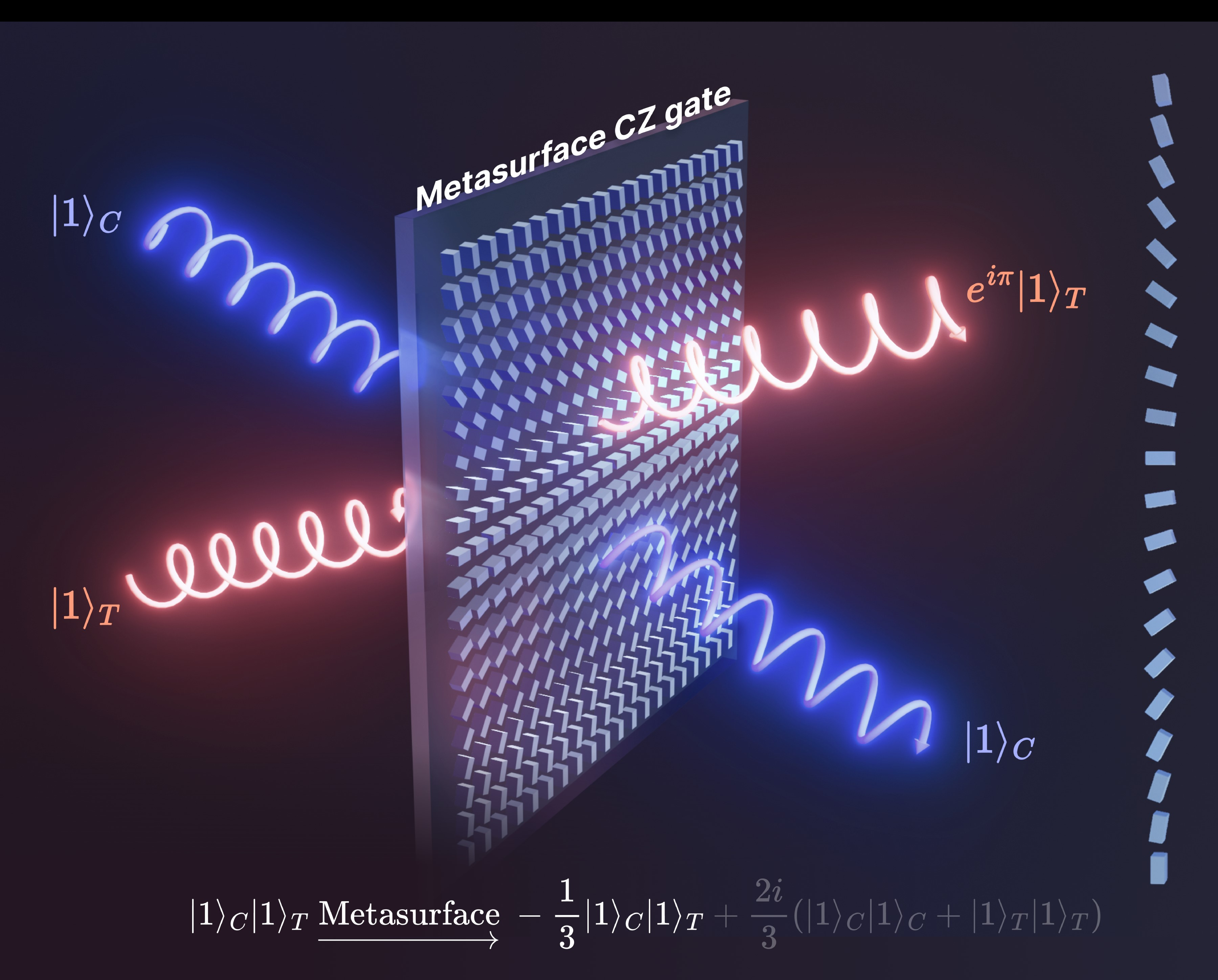}\\
  \caption{Schematic diagram of polarization-encoded quantum CZ gate through single gradient metasurface.
   When the control photon is in the LCP state $|1\rangle_C$ and the target photon in the RCP state $|1\rangle_{T}$, after exiting from the metasurface, there is a relative $\pi$ phase shift between them with a success probability 1/9. For other kinds of input, two output qubits do not have this additional relative phase shift. 
   }\label{FIG1}
\end{figure}

In this Letter, utilizing three of beam splitters in a single gradient metasurface, one quantum CZ gate is first demonstrated[Fig. \ref{FIG1}]. 
Then, choosing another three beam splitters in the same metasurface, the next independent CZ operation can be simultaneously implemented, and so on.
By adding more connected beam splitters into the first CZ operation, cascaded CZ gates can be realized. 
Thus one piece of metasurface can provide novel multiqubit operation, which will be applied to multiqubit entangled state preparation and multiparty quantum swapping.
In addition, taking advantage of the input polarization controlled output path-locking feature, both polarization-encoded and path-encoded CZ gates can be demonstrated on the same metasurface, which further improves the integration level of quantum devices.
Our research may be generalized to realize other integrated quantum tasks, such as multiqubit quantum gates, thus opening an avenue for high-density and multifunction quantum integration on metasurface. 

%%%%%%%%%%%%%%%%%%%%%%%%%%%%%%%Figure 1%%%%%%%%%%%%%%%%%%%%%%%%%%%%%%%%%%%%%%%%%%%%%%%%%
%%%%%%%%%%%%%%%%%%%%%%%%%%%%%%%%%%%%%%%%%%%%%%%%%%%%%%%%%%%%%%%%%%%%%%%%%%%%%%%%

The mechanism of using single metasurface to realize a quantum CZ gate is described as follows.
Based on the principle of linear optics, the circuit of one CZ operation is comprised of three identical beam splitters with 1:2 splitting ratio of power~\cite{SKLM2,SKLM3}.
 It is recently found that polarization-sensitive PB phase metasurfaces can act as parallel beam splitters, i.e., a series of beam splitters with identical, customizable splitting ratios~\cite{MPBS}. 
One can ingeniously choose three among these splitters to demonstrate a CZ operation.
As shown in Fig. 2, two photons are sent to the metasurface with different paths, one of which serves as control qubit and the other as target. 
Through the quantum interference process of parallel beam splitters that three of which are involved,  the single metasurface performs an equivalent CZ operation. 
But owing to the same splitting principle of linear optics as those reported in Refs.~\cite{SKLM2,SKLM3}, its success probability is only $1/9$.

%When the control photon is in an LCP state and the target photon in an RCP state, the metasurface enables quantum interference between the two photons.
%Consequently, the photons exit the metasurface with their polarization states unchanged, but with a relative $\pi$ phase shift accumulated between them, which is the effect of a CZ gate when both qubits are in the state “1”. 
%However, when the control and target photons are in other circular polarization states, the metasurface fails to induce an interaction between the two qubits, preserving the qubits' initial states. 

%%%%%%%%%%%%%%%%%%%%%%
\begin{figure}[tbp]
  \centering
  \includegraphics[width=0.7\textwidth]{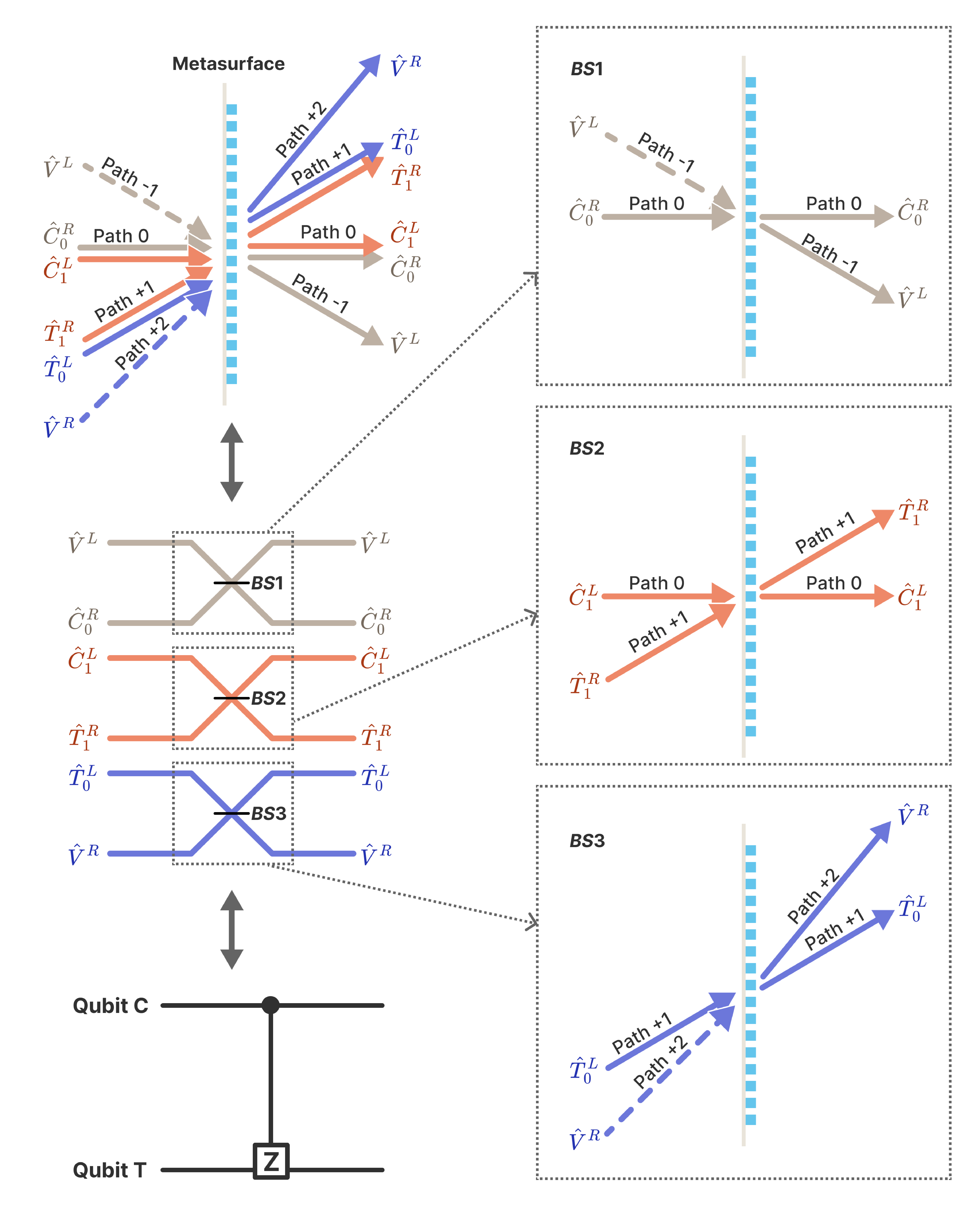}\\
  \caption{Scheme of polarization-encoded quantum CZ gate with the help of parallel beam-splitting on the metasurface.
  %The realization of a quantum CZ gate on a metasurface. (a) Scheme of CZ gate for polarization-encoded qubits with the help of parallel beam-splitting on the metasurface. (b) Diagram of a traditional CZ gate realized by three beam splitters. (c) Equivalent quantum circuit of a CZ gate. (d)-(f) are three independent parallel beam-splitting processes on the same metasurface, each performing the same function as the three beam splitters in (b). These processes (d)-(f) form the configuration of the CZ gate depicted in (a).
}\label{FIG2}
\end{figure}
%%%%%%%%%%%%%%%%%%%%%%%%%%%%%%%%%%%%%%%%%%%%%%%%%%%%%%%%%%%%%%%%%%%%%%%%%%%%%%%%

The gradient metasurface supports a series of diffraction orders or paths that facilitate parallel beam-splitting between polarization modes, but the polarization-path locked relation exists~\cite{MPBS}.
 So when considering the CZ operations,  both polarization and path can be used to encode the qubits.
Here, we let qubits be encoded in left-circular (LCP) and right-circular (RCP) polarization of photons [Tab. I(a)].
As depicted in Fig.~\ref{FIG2}, we select two paths (0,+1) for transmitting qubits C and T, while two auxiliary paths (-1,+2) are on demand to implement the linear optical circuits. 
 Such configuration is equivalent to a traditional CZ gate scheme requiring three beam splitters. 
To see the equivalence, using the principle of parallel beam splitting~\cite{MPBS}, we decompose the six input polarization modes into three pairs of beam-splitting processes [Fig. 2]. Then, the transformation between six modes reads
%%%%%%%%%%
\begin{equation}\label{CZ_Mattrix}
\small
\begin{bmatrix}
\hat{V}^{R}\\
\hat{T}_0^{L}\\
\hat{T}_1^{R}\\
\hat{C}_1^{L}\\
\hat{C}_0^{R}\\
\hat{V}^{L}
\end{bmatrix}_{\rm out}
=\frac{1}{\sqrt{3}}
\begin{bmatrix}
1 & \sqrt{2}i & 0 & 0 & 0 & 0\\
\sqrt{2}i & 1 & 0 & 0 & 0 & 0\\
0 & 0 & 1 & \sqrt{2}i & 0 & 0\\
0 & 0 & \sqrt{2}i & 1 & 0 & 0\\
0 & 0 &0 & 0 & 1 & \sqrt{2}i \\
0 & 0 & 0 & 0 & \sqrt{2}i & 1 
\end{bmatrix}
\begin{bmatrix}
\hat{V}^{R}\\
\hat{T}_0^{L}\\
\hat{T}_1^{R}\\
\hat{C}_1^{L}\\
\hat{C}_0^{R}\\
\hat{V}^{L}
\end{bmatrix}_{in}.
\end{equation}
%%%%%%%%%%%%%
Here, the qubit and auxiliary modes are related to some polarization modes: $[\hat{V}^{R},\hat{T}_0^{L},\hat{T}_1^{R},\hat{C}_1^{L},\hat{C}_0^{R},\hat{V}^{L}]=[\hat{a}_{R}(+2),\hat{a}_{L}(+1),\hat{a}_{R}(+1),\hat{a}_{L}(0),\hat{a}_{R}(0),\hat{a}_{L}(-1)]$, where the annihilation operator $\hat{a}_L(j)$ or $\hat{a}_R(j)$  represents the LCP or RCP photon occupying path $j$~\cite{MPBS}.
The function of  Eq.~(\ref{CZ_Mattrix}) is identical to those integrated waveguides of implementing a  CZ gate~\cite{CHIP1,CHIP3,COMPACT3,COMPACT4}.

%%%%%%%%%%%%%%%%%%%%%%%%%%%%%%%%%%%%%%%%%%%%%%%%
\begin{table}[tbp]\label{Tab1}
    %\centering
    \caption{Characteristic of single CZ gate on metasurface. (a)  Encoding rules of polarization qubits and (b) truth table of  2-qubit input states. }(a)
    \begin{ruledtabular}\label{Tab1}
    \begin{tabular}{lcccccc}
%\hline
        & \multicolumn{2}{c}{Control} & \multicolumn{2}{c}{Target} &\multicolumn{2}{c}{Auxiliary} \\
        \colrule
        Polarization state&$|R\rangle_0$ & $|L\rangle_0$ &$|R\rangle_{+1}$&$|L\rangle_{+1}$&$|R\rangle_{+2}$&$|L\rangle_{-1}$ \\
        Encoded Qubit state&$|0\rangle_C$ & $|1\rangle_C$ & $|1\rangle_T$&$|0\rangle_T$&None&None \\
        Mode Label&$\hat{C}_0^{R}$&$\hat{C}_1^{L}$&$\hat{T}_1^{R}$&$\hat{T}_0^{L}$&$\hat{V}^{R}$&$\hat{V}^{L}$\\
    \end{tabular}
    \end{ruledtabular}
\vspace{5pt}
(b)
\begin{ruledtabular}
    \begin{tabular}{lrc}
        Input & Output~ & Success probability\\
        \colrule
        $|0\rangle_C|0\rangle_T$ &$|0\rangle_C|0\rangle_T$ &1/9\\
        $|0\rangle_C|1\rangle_T$ &$|0\rangle_C|1\rangle_T$ &1/9\\
        $|1\rangle_C|0\rangle_T$ &$|1\rangle_C|0\rangle_T$ &1/9\\
        $|1\rangle_C|1\rangle_T$ &$-|1\rangle_C|1\rangle_T$ &1/9
    \end{tabular}
    \end{ruledtabular}
\end{table}
%%%%%%%%%%%%%%%%%%%%%%%%%%%%%%%%%%%%%%%%%%%%%%%%%%

With above setup, the truth table of implementing  CZ operation is shown in Table~\ref{Tab1}(b). 
When the input state is $|1\rangle_C|1\rangle_T$, i.e., $\hat{C}_1^L$,$\hat{T}_1^R$ modes are occupied by one photon respectively, then according to Eq.~(\ref{CZ_Mattrix}),  the output state becomes,
%%%%%%%%%%%%%%%%%%%%%%%%%%%%%%%%%%%%%%%%%%%%%%%%%%%%%%%%%%%%%
\begin{equation}\label{CZout_11}
\small
|1\rangle_C|1\rangle_T~\underrightarrow{\text{~~Metasurface~~}}~-\frac{1}{3}|1\rangle_C|1\rangle_T+\frac{2i}{3}(|1\rangle_C|1\rangle_C+|1\rangle_T|1\rangle_T),
\end{equation}
%%%%%%%%%%%%%%%%%%%%%%%%%%%%%%%%%%%%%%%%%%%%%%%%%%%%%%%%%%%%%%
with a success probability of 1/9 to obtain the state $-|1\rangle_C|1\rangle_T$. 
%Also, there is a probability of 8/9 that the CZ gate operation will fail because qubits are missing (two photon occupy the same qubit mode is logically meaningless). 
The quantum interference originating from the parallel beam-splitting process of the metasurface 
induces an additional $\pi$ phase shift or negative sign in Eq.~(\ref{CZout_11}), which guarantees the realization of a CZ operation.
While for other kinds of input states, the corresponding state transformations are shown in Table~\ref{Tab1}(b). These results confirm that a CZ gate has been implemented on single metasurface with a success probability of 1/9, consistent with previously reported CZ gates~\cite{SKLM3}.

For a single quantum CZ gate, only two adjacent diffraction orders are selected to transmit two qubits.
In fact, under the paraxial approximation, the transformation matrix for any two adjacent diffraction orders has the same form as the $2\times2$ matrix block in Eq.~(\ref{CZ_Mattrix})~\cite{MPBS}. 
If we choose another two adjacent diffraction orders as transmitting qubits C and T, then on the same metasurface, a new CZ operation can simultaneously exist.  
Therefore, as long as the modes selected for these sets of paths do not interfere with each other, these CZ gates can operate independently. For example, except for paths (0, +1) shown in Fig.~2, we can use paths (-2, -3) to construct another independent CZ gate. In a similar way, more CZ gates are available in the same single metasurface.

%Secondly, more than one CZ gate is available on the same metasurface.It is the direct sum of these matrix blocks that constitutes the entire parallel beam-splitting response of the metasurface~\cite{MPBS}. Therefore, we can choose other input paths to transmit the control and target qubits, as long as the corresponding diffraction orders are adjacent. By adding two auxiliary modes, we obtain the same transformation shown in Eq.~(\ref{CZ_Mattrix}), thus supporting a new CZ gate. 

%%%%%%%%%%%%%%%%%%%%%%
\begin{figure}[tbp]
  \centering
  \includegraphics[width=0.7\textwidth]{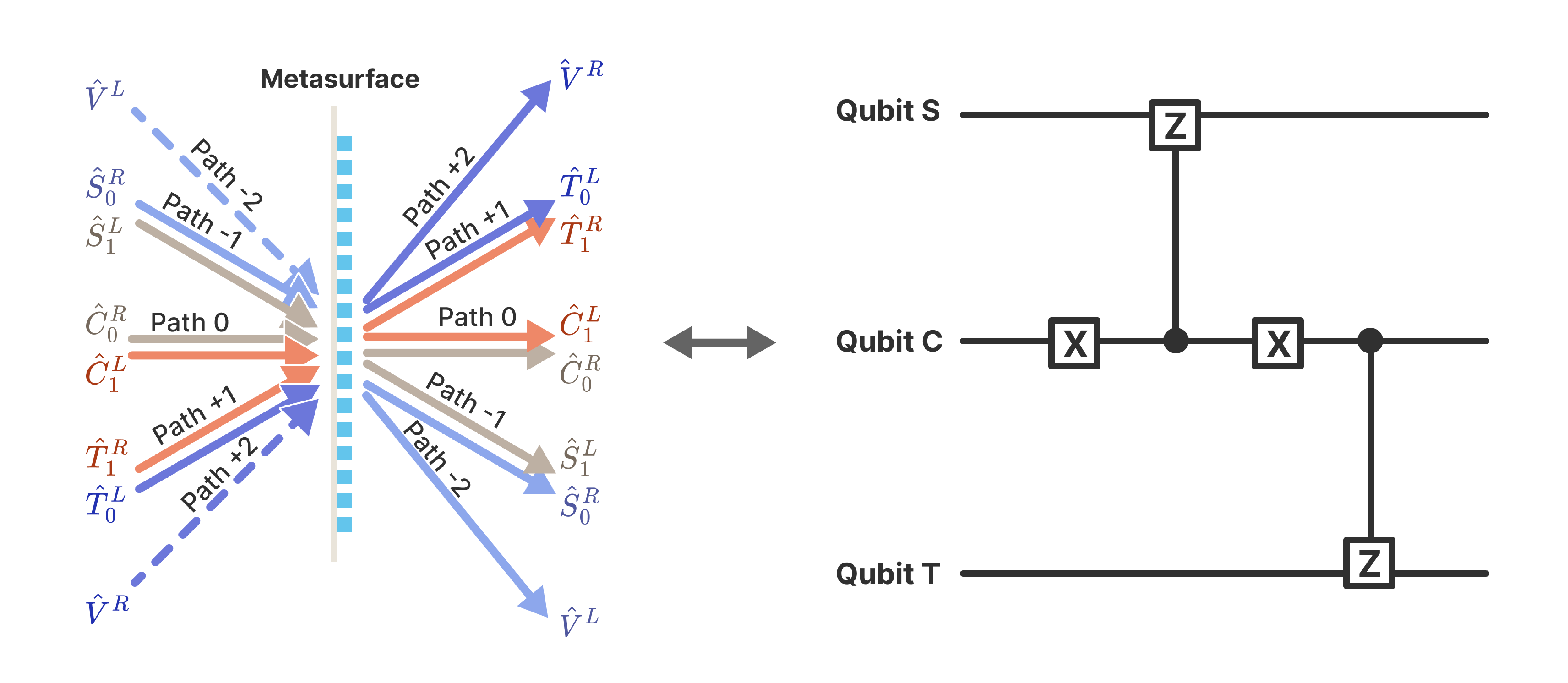}\\
  \caption{Scheme of a cascaded quantum CZ gate with the help of parallel beam-splitting on the metasurface.
  %The realization of two cascaded quantum CZ gates on a metasurface. (a) Scheme of cascaded CZ gates for polarization-encoded qubits with the help of parallel beam-splitting on the metasurface. (b) Equivalent quantum circuit of two cascaded quantum CZ gates.
}\label{FIG3}
\end{figure}
%%%%%%%%%%%%%%%%%%%%%%%%%%%%%%%%%%%%%%%%%%%%%%%%%%%%%%%%%%%%%%%%%%%%%%%%%%%%%%%%

%Thirdly, CZ gates on the metasurface can be cascaded to build a multiqubit quantum circuit. 
When two CZ gates supported by single metasurface share a common qubit, they are no longer independent and become cascaded. 
Figure 3(a) illustrates this cascading configuration, where paths (0, +1) form one CZ gate, and paths (-1, 0) form another, both of which share the path 0 (qubit C). 
For simplicity, the paths (0, +1) continue to transmit qubits T and C, keeping the same encoding rule as in Fig.~\ref{FIG2} and Table~\ref{Tab1}, while the third qubit, denoted as S, is encoded as in Table ~\ref{Tab2}(a). Then, the transformation relation of input-output qubit  reads
%%%%%%%%%%
\begin{equation}\label{CZ_Mattrix2}
\small
\begin{bmatrix}
\hat{V}^{R}\\
\hat{T}_0^{L}\\
\hat{T}_1^{R}\\
\hat{C}_1^{L}\\
\hat{C}_0^{R}\\
\hat{S}_1^{L}\\
\hat{S}_0^{R}\\
\hat{V}^{L}
\end{bmatrix}_{\rm out}
=\frac{1}{\sqrt{3}}
\begin{bmatrix}
1 & \sqrt{2}i & 0 & 0 & 0 & 0 & 0 & 0\\
\sqrt{2}i & 1 & 0 & 0 & 0 & 0 & 0 & 0\\
0 & 0 & 1 & \sqrt{2}i & 0 & 0 & 0 & 0\\
0 & 0 & \sqrt{2}i & 1 & 0 & 0 & 0 & 0\\
0 & 0 &0 & 0 & 1 & \sqrt{2}i & 0 & 0\\
0 & 0 & 0 & 0 & \sqrt{2}i & 1 & 0 & 0\\
0 & 0 &0 & 0  & 0 & 0& 1 & \sqrt{2}i\\
0 & 0 & 0 & 0  & 0 & 0& \sqrt{2}i & 1
\end{bmatrix}
\begin{bmatrix}
\hat{V}^{R}\\
\hat{T}_0^{L}\\
\hat{T}_1^{R}\\
\hat{C}_1^{L}\\
\hat{C}_0^{R}\\
\hat{S}_1^{L}\\
\hat{S}_0^{R}\\
\hat{V}^{L}
\end{bmatrix}_{in},
\end{equation}
%%%%%%%%%%%%%
where the qubit modes and auxiliary modes are related to some different polarization modes, which are $[\hat{V}^{R},\hat{T}_0^{L},\hat{T}_1^{R},\hat{C}_1^{L},\hat{C}_0^{R},\hat{S}_1^{L},\hat{S}_0^{R},\hat{V}^{L}]=[\hat{a}_{R}(+2),\hat{a}_{L}(+1),\hat{a}_{R}(+1),\hat{a}_{L}(0),\hat{a}_{R}(0),\hat{a}_{L}(-1),$ $\hat{a}_{R}(-1),\hat{a}_{L}(-2)]$.

\begin{table}[htbp]
\centering
 \caption{Characteristic of a two-cascaded CZ gate on metasurface. (a) Encoding rules of signal polarization qubit and (b)  truth table of 3-qubit input states.}\label{Tab2}
(a)
\begin{ruledtabular}
\centering
\begin{tabular}{lcccc}
%\toprule[1.5pt]
 & \multicolumn{2}{c}{Signal} &\multicolumn{2}{c}{Auxiliary} \\ \colrule
 Polarization state&$|R\rangle_{-1}$&$|L\rangle_{-1}$&$|R\rangle_{+2}$&$|L\rangle_{-2}$ \\
        Encoded Qubit state& $|0\rangle_S$&$|1\rangle_S$&None&None \\
        Mode Label&$\hat{S}_0^{R}$&$\hat{S}_1^{L}$&$\hat{V}^{R}$&$\hat{V}^{L}$ \\ %\bottomrule[1.5pt]
 \end{tabular}
\end{ruledtabular}
\vspace{5pt}
(b)
\begin{ruledtabular}
\centering
\begin{tabular}{lrcccccc} %\toprule[1.5pt]
 \multicolumn{2}{c}{S\&T in standard basis} &~&\multicolumn{2}{c}{S\&T in Hadamard basis} &~\\
\cline{1-2}
\cline{4-5}
Input & Output~~~~~ &~&Input & Output & Success probability\\
        $|0\rangle_C|0\rangle_S|0\rangle_T$ &$|0\rangle_C|0\rangle_S|0\rangle_T$ &~&$|0\rangle_C|+\rangle_S|+\rangle_T$ &$|0\rangle_C|-\rangle_S|+\rangle_T$ &1/27\\
        $|0\rangle_C|0\rangle_S|1\rangle_T$ &$|0\rangle_C|0\rangle_S|1\rangle_T$ &~&$|0\rangle_C|+\rangle_S|-\rangle_T$ &$|0\rangle_C|-\rangle_S|-\rangle_T$ &1/27\\
        $|0\rangle_C|1\rangle_S|0\rangle_T$ &$-|0\rangle_C|1\rangle_S|0\rangle_T$ &~&$|0\rangle_C|-\rangle_S|+\rangle_T$ &$|0\rangle_C|+\rangle_S|+\rangle_T$ &1/27\\
        $|0\rangle_C|1\rangle_S|1\rangle_T$ &$-|0\rangle_C|1\rangle_S|1\rangle_T$ &~&$|0\rangle_C|-\rangle_S|-\rangle_T$ &$|0\rangle_C|+\rangle_S|-\rangle_T$ &1/27\\
        $|1\rangle_C|0\rangle_S|0\rangle_T$ &$|1\rangle_C|0\rangle_S|0\rangle_T$ &~&$|1\rangle_C|+\rangle_S|+\rangle_T$ &$|1\rangle_C|+\rangle_S|-\rangle_T$ &1/27\\
        $|1\rangle_C|0\rangle_S|1\rangle_T$ &$-|1\rangle_C|0\rangle_S|1\rangle_T$ &~&$|1\rangle_C|+\rangle_S|-\rangle_T$ &$|1\rangle_C|+\rangle_S|+\rangle_T$ &1/27\\
        $|1\rangle_C|1\rangle_S|0\rangle_T$ &$|1\rangle_C|1\rangle_S|0\rangle_T$ &~&$|1\rangle_C|-\rangle_S|+\rangle_T$ &$|1\rangle_C|-\rangle_S|-\rangle_T$ &1/27\\
        $|1\rangle_C|1\rangle_S|1\rangle_T$ &$-|1\rangle_C|1\rangle_S|1\rangle_T$ &~&$|1\rangle_C|-\rangle_S|-\rangle_T$ &$|1\rangle_C|-\rangle_S|+\rangle_T$ &1/27
 \\ %\bottomrule[1.5pt]
 \end{tabular}
\end{ruledtabular}
\end{table}
%%%%%%%%%%%%%%%%%%%%%%%%%%%%%%%%%%%%%%%%%%%%%%%%%%

Employing similar methods, according to Eq. (3), we analyze the transformations for different qubit input states. 
The results in Table~\ref{Tab2}(b) indicate that the operation through the metasurface corresponds to the quantum circuit of two sequential CZ operations shown in Fig.~\ref{FIG3} with a success probability of 1/27. 
The additional X gates in this circuit come from different encoding rules of two CZ gates.
%{\color{red}From the truth table, one can see that, if signal qubit is $|0\rangle_S$, those previous CZ operations are not changed. If signal qubit is $|1\rangle_S$, for $|0\rangle_C|0\rangle_T$ and $|0\rangle_C|1\rangle_T$, there is a $\pi$ phase shift in the output state; while for  $|1\rangle_C|0\rangle_T$ and $|1\rangle_C|1\rangle_T$, they still keep previous CZ operation. So in this cascaded CZ gate, the operation for  the states $|0\rangle_C|0\rangle_T$ and $|0\rangle_C|1\rangle_T$ is executed with the success probability of 1/27.}
\textcolor{black}{From the equivalent quantum circuit and the truth tables, it is clear that qubit C controls both qubits S and T, but with different control conditions. Under the standard basis, when qubit C is $|0\rangle_C$, qubit T is unaffected, and qubit S acquires a phase shift (negative sign) when it is in state $|1\rangle_S$. Conversely, when qubit C is $|1\rangle_C$, qubit S remains unchanged, and qubit T acquires a phase shift when it is $|1\rangle_T$. This quantum logical function is more evident when qubits S and T are in the Hadamard basis $|\pm\rangle_{S,T}=(|0\rangle_{S,T}\pm|1\rangle_{S,T})/\sqrt{2}$, where qubit S flips when qubit C is $|0\rangle_C$, and qubit T flips when qubit C is $|1\rangle_C$.}
Repeating this process, in principle, one metasurface can cascade more CZ gates, which is equivalent to performing multiple sequential CZ operations. But the number of cascading gates is limited by the diffraction orders that the metasurface can support.
Thus, the parallel beam-splitting capability of single metasurface not only enables the implementation of  CZ gates, but also allows for the direct construction of cascaded quantum circuits.
A single CZ gate can entangle two qubits~\cite{CNOT_QIP3}, while a quantum circuit with cascaded CZ gates can establish multiqubit entanglement by sequentially entangling qubit pairs.
For instance, if the initial qubit state is set as $|+\rangle_S|+\rangle_C|+\rangle_T$, where $|\pm\rangle=(|0\rangle\pm|1\rangle)/\sqrt{2}$ (corresponding to all three photons in horizontal linear polarization), then we can use the metasurface configuration depicted in Fig.~\ref{FIG3} to prepare the entangled state.
According to Eq. (3), if the quantum circuit operation is successfully executed, the output state will evolve to $(|+\rangle_S|1\rangle_C|-\rangle_T+|-\rangle_S|0\rangle_C|+\rangle_T)/\sqrt{2}$, which is a three-qubit GHZ entangled state. 
The cascaded CZ operations on the metasurface provide a convenient method for preparing multiqubit entangled states, which will be also utilized for multiqubit entanglement swapping~\cite{SWAP1,SWAP2}.

%%%%%%%%%%%%%%%%%%%%%%
\begin{figure}[tbp]
  \centering
  \includegraphics[width=0.7\textwidth]{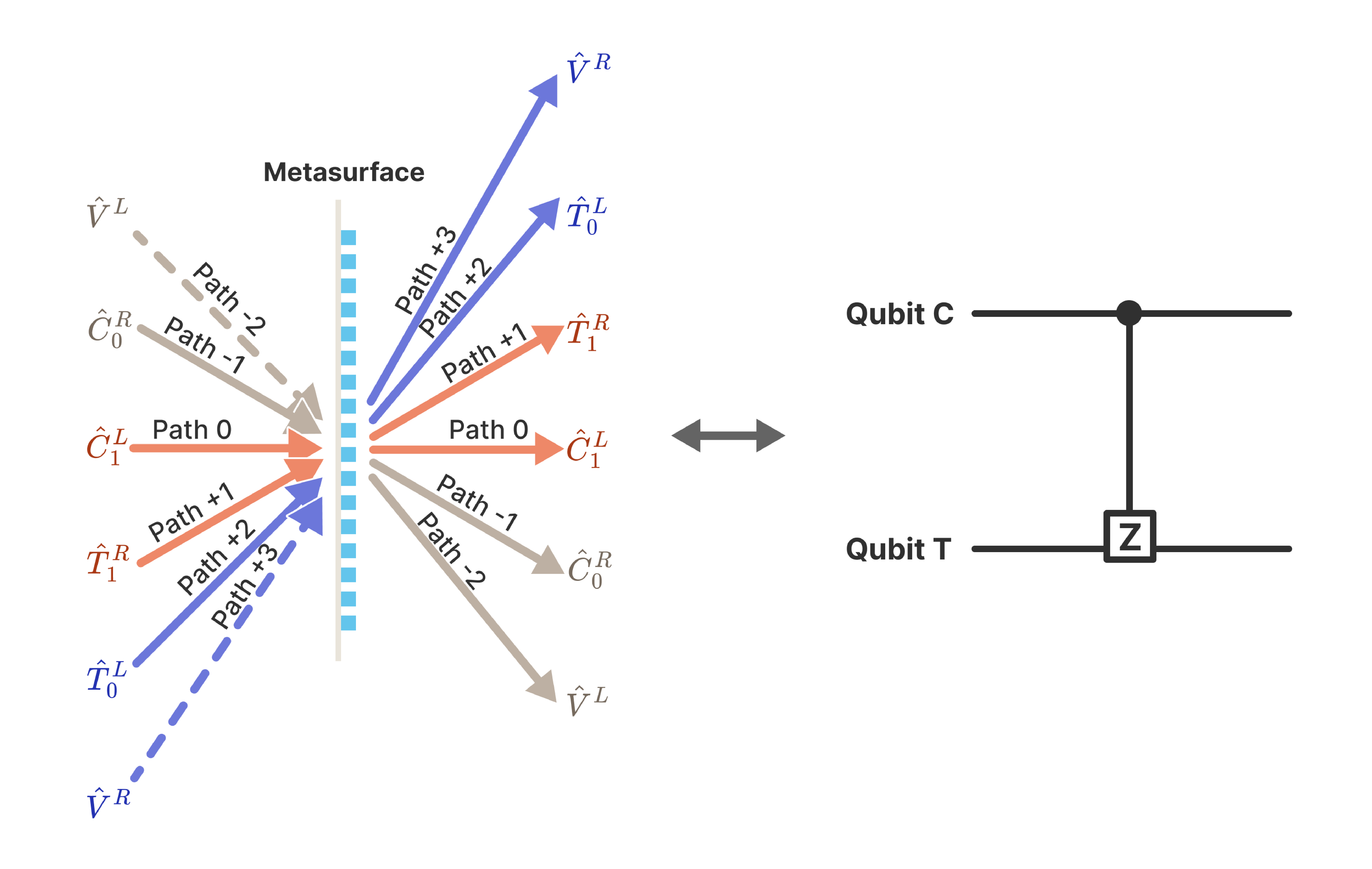}\\
  \caption{Scheme of path-encoded quantum CZ gate with the help of parallel beam-splitting on the metasurface.
}\label{FIG4}
\end{figure}
%%%%%%%%%%%%%%%%%%%%%%%%%%%%%%%%%%%%%%%%%%%%%%%%%%%%%%%%%%%%%%%%%%%%%%%%%%%%%%%%

\textcolor{black}{We note that the parallel beam-splitting feature of metasurfaces also enables the implementation of path-encoded quantum CZ gates. 
As illustrated in Fig.~\ref{FIG4}, a single qubit is carried by two distinct paths. 
Under this configuration, there are still three parallel beam-splitting processes, whose overall beam-splitting transformation matrix remains consistent with Eq.~(\ref{CZ_Mattrix}). 
Different from the polarization encoding scheme, the mode correspondence here is $[\hat{V}^{R},\hat{T}_0^{L},\hat{T}_1^{R},\hat{C}_1^{L},\hat{C}_0^{R},\hat{V}^{L}]=[\hat{a}_{R}(+3),\hat{a}_{L}(+2),\hat{a}_{R}(+1),\hat{a}_{L}(0),\hat{a}_{R}(-1),\hat{a}_{L}(-2)]$, indicating three beam-splitting processes. 
%In contrast, in the polarization encoding scheme, the two selected adjacent parallel beam-splitting processes share a common path. 
%It is noteworthy that the CZ gates under both encoding schemes are implemented on the same metasurface, and different mode selection methods can facilitate the switch between the two schemes. Typically, switching between the two encoding methods in previous schemes required additional devices, making it challenging to realize both on the same device simultaneously. 
Then, using the same truth table as that in Tab. 1, a CZ operation can be performed.
Thus, metasurfaces offer great flexibility in the quantum logical operations.
If combining both path and polarization encoding methods together, it is possible to realize the high-dimensional qudit logic gates on a single metasurface.}

We will now briefly explore the experimental feasibility of our proposed scheme. A high-efficiency metasurface can be fabricated using electron beam lithography with high-index dielectric materials, such as amorphous silicon~\cite{MSBS1}, silicon nitride~\cite{MSBS5} for near-infrared wavelengths or titanium dioxide~\cite{MSBS2} for visible light. The experimental diffraction efficiency is expected to reach approximately $40\%\sim70\%$ or higher~\cite{MQS,MSBS1,MSBS2,MSBS5}, which will slightly reduce the success probability of the quantum gate. The gate fidelity is primarily affected by non-ideal 1:2 splitting ratios and imperfect circular polarization conversion. However, thanks to current micro-/nanofabrication technology, the error in splitting ratios can be kept within 5\%, and polarization conversion efficiency can approach unity~\cite{MSBS2,MSBS5}. Additionally, the oblique incidence scheme in this work will also influence gate fidelity due to the gradual change in splitting ratios as the incident angle increases. This issue can be mitigated by operating under paraxial conditions, ensuring that the splitting ratio error remains within 5\%. Within the aforementioned error range, the gate fidelity can exceed 90\%~\cite{CHIP1,COMPACT2,COMPACT3,COMPACT4}. Thus, our scheme is highly likely to be realized experimentally.

%Secondly, our findings suggest that, in addition to reduce device dimensions~\cite{COMPACT1,COMPACT2,COMPACT3,COMPACT4}, enhancing parallel and multifunctional operation capabilities on a single micro-nano device can also improve integration density. Thirdly, we have shown a novel quantum manipulation capability of metasurfaces. Unlike previous applications that leverage metasurfaces for parallel operations on a single quantum source~\cite{MQM1,MQM3,MQM4,MQS}, we explore the ability to perform parallel or correlated manipulation of multiple quantum light sources or states. These results also provide insights for further research into multifunctional quantum manipulation on metasurfaces. For instance, by integrating multilayer metasurface, more complex multiqubit gates, quantum circuits, and corresponding multi-degree-of-freedom quantum state control and preparation can be realized.
{\color{black}
To summarize, 
we have demonstrated a quantum CZ gate, several independent CZ gates, and a cascaded CZ gate on a single gradient metasurface through the principle of parallel beam-splitting.
In these CZ operations, both polarization and path qubits can be encoded with the same metasurface, which further improves the integration level of quantum devices.
Though we have demonstrated fundamental two-qubit logical gates, based on the same principle, multi-qubit logical operations could be executed on the single metasurface, which will open the research of metasurface-based multifunctional quantum manipulation and quantum integration. 
 Thus, this work paves the way for high-density, multifunctional quantum integration on metasurfaces, with potential applications in on-chip quantum optics and quantum information processing.
}
\acknowledgments
\textit{Acknowledgments.} This work is supported by the National Natural Science Foundation of China under Grants No. 11974032, No. 12161141010 and No. T2325022, and by the Innovation Program for Quantum Science and Technology under Grant No. 2021ZD0301500.

%\nocite{*}

%\bibliography{EV-whole}

\begin{thebibliography}{99}
%%%%MSR: Review of metasurface concepts and light field manipulation
\bibitem{MSR1}	N. Yu, P. Genevet, M. A. Kats, F. Aieta, J.- P. Tetienne, F. Capasso, and Z. Gaburro, {\it Light Propagation with Phase Discontinuities: Generalized Laws of Reflection and Refraction}, Science \textbf{334}, 333 (2011).

\bibitem{MSR2}	H.- T. Chen, A. J. Taylor, and N. Yu, {\it A Review of Metasurfaces: Physics and Applications}, Rep. Prog. Phys. \textbf{79}, 076401 (2016).

\bibitem{MSR3}	G. Li, S. Zhang, and T. Zentgraf, {\it Nonlinear Photonic Metasurfaces}, Nat. Rev. Mater. \textbf{2}, 17010 (2017).

\bibitem{MSR4}	C.- W. Qiu, T. Zhang, G. Hu, and Y. Kivshar, {\it Quo Vadis, Metasurfaces?}, Nano Lett. \textbf{21}, 5461 (2021).

\bibitem{MSR5}	A. H. Dorrah and F. Capasso, {\it Tunable Structured Light with Flat Optics}, Science \textbf{376}, 367 (2022).
%%%%MQR: Review of quantum photonics in metasurface
\bibitem{MQR1}	A. S. Solntsev, G. S. Agarwal, and Y. S. Kivshar, {\it Metasurfaces for Quantum Photonics}, Nat. Photonics \textbf{15}, 327 (2021).

\bibitem{MQR2}	K. Wang, M. Chekhova, and Y. Kivshar, Metasurfaces for quantum technologies, Phys. Today \textbf{75}, 8, 38 (2022).

%%%%MQP: High-Dimensional/Multiphoton quantum states preparation with metasurface
\bibitem{MQP1}	L. Li, Z. Liu, X. Ren, S. Wang, V.- C. Su, M.- K. Chen, C. H. Chu, H. Y. Kuo, B. Liu, W. Zang, G. Guo, L. Zhang, Z. Wang, S. Zhu, and D. P. Tsai, {\it Metalens-Array-Based High-Dimensional and Multiphoton Quantum Source}, Science \textbf{368}, 1487 (2020).

\bibitem{MQP2}	W. J. M. Kort-Kamp, A. K. Azad, and D. A. R. Dalvit, {\it Space-Time Quantum Metasurfaces}, Phys. Rev. Lett. \textbf{127}, 043603 (2021).

\bibitem{MQP3}	T. Santiago-Cruz, S. D. Gennaro, O. Mitrofanov, S. Addamane, J. Reno, I. Brener, and M. V. Chekhova, {\it Resonant Metasurfaces for Generating Complex Quantum States}, Science \textbf{377}, 991 (2022).

%%%%MQM: Quantum states manipulation with metasurface
\bibitem{MQM1}	T. Stav, A. Faerman, E. Maguid, D. Oren, V. Kleiner, E. Hasman, and M. Segev, {\it Quantum Entanglement of the Spin and Orbital Angular Momentum of Photons Using Metamaterials}, Science \textbf{361}, 1101 (2018).

\bibitem{MQM2}	Q. Li, W. Bao, Z. Nie, Y. Xia, Y. Xue, Y. Wang, S. Yang, and X. Zhang, {\it A Non-Unitary Metasurface Enables Continuous Control of Quantum Photon-Photon Interactions from Bosonic to Fermionic}, Nat. Photonics \textbf{15}, 267 (2021).

\bibitem{MQM3}	D. Zhang, Y. Chen, S. Gong, W. Wu, W. Cai, M. Ren, X. Ren, S. Zhang, G. Guo, and J. Xu, {\it All-Optical Modulation of Quantum States by Nonlinear Metasurface}, Light-Sci. Appl. \textbf{11}, 58 (2022).

\bibitem{MQM4}	Y.- J. Gao, Z. Wang, Y. Jiang, R.- W. Peng, Z.- Y. Wang, D.- X. Qi, R.- H. Fan, W.- J. Tang, and M. Wang, {\it Multichannel Distribution and Transformation of Entangled Photons with Dielectric Metasurfaces}, Phys. Rev. Lett. \textbf{129}, 023601 (2022).

%%%%MQBS: Metasurface quantum tomography
\bibitem{MQT}	K. Wang, J. G. Titchener, S. S. Kruk, L. Xu, H.- P. Chung, M. Parry, I. I. Kravchenko, Y.- H. Chen, A. S. Solntsev, Y. S. Kivshar, D. N. Neshev, and A. A. Sukhorukov, {\it Quantum Metasurface for Multiphoton Interference and State Reconstruction}, Science \textbf{361}, 1104 (2018).

%%%%MQI: Quantum imaging with metasurface
\bibitem{MQI1}	J. Zhou, S. Liu, H. Qian, Y. Li, H. Luo, S. Wen, Z. Zhou, G. Guo, B. Shi, and Z. Liu, {\it Metasurface Enabled Quantum Edge Detection}, Sci. Adv. \textbf{6}, eabc4385 (2020).

\bibitem{MQI2}	A. Vega, T. Pertsch, F. Setzpfandt, and A. A. Sukhorukov, {\it Metasurface-Assisted Quantum Ghost Discrimination of Polarization Objects}, Phys. Rev. Appl. \textbf{16}, 064032 (2021).

\bibitem{MQI3}	T. K. Yung, H. Liang, J. Xi, W. Y. Tam, and J. Li, {\it Jones-Matrix Imaging Based on Two-Photon Interference}, Nanophotonics \textbf{12}, 579 (2023).

%%%%MQG: Quantum sensing with metasurface
\bibitem{MQS}	P. Georgi, M. Massaro, K.- H. Luo, B. Sain, N. Montaut, H. Herrmann, T. Weiss, G. Li, C. Silberhorn, and T. Zentgraf, {\it Metasurface Interferometry toward Quantum Sensors}, Light-Sci. Appl. \textbf{8}, 70 (2019).

%%%%MSBS: Metasurface polarization beam splitters
\bibitem{MSBS1}	A. Arbabi, Y. Horie, M. Bagheri, and A. Faraon, {\it Dielectric Metasurfaces for Complete Control of Phase and Polarization with Subwavelength Spatial Resolution and High Transmission}, Nat. Nanotechnol. \textbf{10}, 937 (2015).

\bibitem{MSBS2}	J. P. Balthasar Mueller, N. A. Rubin, R. C. Devlin, B. Groever, and F. Capasso, {\it Metasurface Polarization Optics: Independent Phase Control of Arbitrary Orthogonal States of Polarization}, Phys. Rev. Lett. \textbf{118}, 113901 (2017).

\bibitem{MSBS3}	V. S. Asadchy, A. D\'{i}az-Rubio, S. N. Tcvetkova, D.- H. Kwon, A. Elsakka, M. Albooyeh, and S. A. Tretyakov, {\it Flat Engineered Multichannel Reflectors}, Phys. Rev. X \textbf{7}, 031046 (2017).

\bibitem{MSBS4}	Y.- J. Gao, X. Xiong, Z. Wang, F. Chen, R.- W. Peng, and M. Wang, {\it Simultaneous Generation of Arbitrary Assembly of Polarization States with Geometrical-Scaling-Induced Phase Modulation}, Phys. Rev. X \textbf{10}, 031035 (2020).

\bibitem{MSBS5}	M. Jin, X. Zhang, X. Liu, C. Liang, J. Liu, Z. Hu, K. Li, G. Wang, J. Yang, L. Zhu, and G. Li, {\it A Centimeter-Scale Dielectric Metasurface for the Generation of Cold Atoms}, Nano Lett. \textbf{23}, 4008-4013 (2023).

%%%%MPBS: Metasurface parallel beam-splitting
\bibitem{MPBS} Q. Liu, X. Liu, Y. Tian, Z. Tian, G. Li, X.-F. Ren, Q. Gong, and Y. Gu, {\it A Parallel Beam Splitting Based on Gradient Metasurface: Preparation and Fusion of Quantum Entanglement}, arXiv:2403.08233 [physics.optics] (2024).



%%%%CNOT_QIP: Importance of CNOT/CZ gate in quantum information process
\bibitem{CNOT_QIP1} P. Kok, W. J. Munro, K. Nemoto, T. C. Ralph, J. P. Dowling, and G. J. Milburn, {\it Linear Optical Quantum Computing with Photonic Qubits}, Rev. Mod. Phys. \textbf{79}, 135 (2007).

\bibitem{CNOT_QIP2} J. Wang, F. Sciarrino, A. Laing, and M. G. Thompson, {\it Integrated photonic quantum technologies}, Nat. Photonics \textbf{14}, 273–284 (2020)

\bibitem{CNOT_QIP3} M. A. Nielsen and I. L. Chuang, {\it Quantum Computation and Quantum Information}, 10th ed. (Cambridge University Press, New York, 2011).

%%%%KLM: Linear optics scheme for quantum computation
\bibitem{KLM} E. Knill, R. Laflamme, and G. J. Milburn, {\it A Scheme for Efficient Quantum Computation with Linear Optics}, Nature \textbf{409}, 46 (2001).

%%%%SKLM: Simplify and realization of KLM CNOT/CZ
\bibitem{SKLM1} T. B. Pittman, B. C. Jacobs, and J. D. Franson, {\it Probabilistic Quantum Logic Operations Using Polarizing Beam Splitters}, Phys. Rev. A \textbf{64}, 062311 (2001).

\bibitem{SKLM2} T. C. Ralph, N. K. Langford, T. B. Bell, and A. G. White, {\it Linear Optical Controlled-Not Gate in the Coincidence Basis}, Phys. Rev. A \textbf{65}, 062324 (2002). 

\bibitem{SKLM3} H. F. Hofmann and S. Takeuchi, {\it Quantum Phase Gate for Photonic Qubits Using Only Beam Splitters and Postselection}, Phys. Rev. A \textbf{66}, 024308 (2002).

\bibitem{SKLM4} J. L. O'Brien, G. J. Pryde, A. G. White, T. C. Ralph, and D. Branning, {\it Demonstration of an All-Optical Quantum Controlled-Not Gate}, Nature \textbf{426}, 264 (2003).

\bibitem{SKLM5} S. Gasparoni, J.-W. Pan, P. Walther, T. Rudolph, and A. Zeilinger, {\it Realization of a Photonic Controlled-Not Gate Sufficient for Quantum Computation}, Phys. Rev. Lett. \textbf{93}, 020504 (2004).

\bibitem{SKLM6} N. K. Langford, T. J. Weinhold, R. Prevedel, K. J. Resch, A. Gilchrist, J. L. O’Brien, G. J. Pryde, and A. G. White, {\it Demonstration of a Simple Entangling Optical Gate and Its Use in Bell-State Analysis}, Phys. Rev. Lett. \textbf{95}, 210504 (2005).

\bibitem{SKLM7} N. Kiesel, C. Schmid, U. Weber, R. Ursin, and H. Weinfurter, {\it Linear Optics Controlled-Phase Gate Made Simple}, Phys. Rev. Lett. \textbf{95}, 210505 (2005).

\bibitem{SKLM8} R. Okamoto, H. F. Hofmann, S. Takeuchi, and K. Sasaki, {\it Demonstration of an Optical Quantum Controlled-Not Gate without Path Interference}, Phys. Rev. Lett. \textbf{95}, 210506 (2005).

%%%%CHIP: Path and polarization encode CZ/CNOT gate on chip
\bibitem{CHIP1} A. Politi, M. J. Cryan, J. G. Rarity, S. Yu, and J. L. O'Brien, {\it Silica-on-Silicon Waveguide Quantum Circuits}, Science \textbf{320}, 646 (2008).

\bibitem{CHIP2} T. Meany, D. N. Biggerstaff, M. A. Broome, A. Fedrizzi, M. Delanty, M. J. Steel, A. Gilchrist, G. D. Marshall, A. G. White, and M. J. Withford, {\it Engineering Integrated Photonics for Heralded Quantum Gates}, Sci. Rep. \textbf{6}, 25126 (2016).

\bibitem{CHIP3} Q. Zhang, M. Li, Y. Chen, X. Ren, R. Osellame, Q. Gong, and Y. Li, {\it Femtosecond Laser Direct Writing of an Integrated Path-Encoded Cnot Quantum Gate}, Opt. Mater. Express \textbf{9}, 2318 (2019).

\bibitem{CHIP4} A. Crespi, R. Ramponi, R. Osellame, L. Sansoni, I. Bongioanni, F. Sciarrino, G. Vallone, and P. Mataloni, {\it Integrated Photonic Quantum Gates for Polarization Qubits}, Nat. Commun. \textbf{2}, 566 (2011).

\bibitem{CHIP5} J. Zeuner, A. N. Sharma, M. Tillmann, R. Heilmann, M. Gräfe, A. Moqanaki, A. Szameit, and P. Walther, {\it Integrated-Optics Heralded Controlled-Not Gate for Polarization-Encoded Qubits}, npj Quantum Inform. \textbf{4}, 13 (2018).

%%%%COMPACT: Compact CNOT gate on chip
\bibitem{COMPACT1} S. M. Wang, Q. Q. Cheng, Y. X. Gong, P. Xu, C. Sun, L. Li, T. Li, and S. N. Zhu, {\it A $14\times14~\mu m^2$ Footprint Polarization-Encoded Quantum Controlled-NOT Gate Based on Hybrid Waveguide}, Nat. Commun. \textbf{7}, 11490 (2016).

\bibitem{COMPACT2} L.-T. Feng, M. Zhang, X. Xiong, D. Liu, Y.-J. Cheng, F.-M. Jing, X.-Z. Qi, Y. Chen, D.-Y. He, G.-P. Guo, G.-C. Guo, D.-X. Dai, and X.-F. Ren, {\it Transverse Mode-Encoded Quantum Gate on a Silicon Photonic Chip}, Phys. Rev. Lett. \textbf{128}, 060501 (2022).

\bibitem{COMPACT3} M. Zhang, L. Feng, M. Li, Y. Chen, L. Zhang, D. He, G. Guo, G. Guo, X. Ren, and D. Dai, {\it Supercompact Photonic Quantum Logic Gate on a Silicon Chip}, Phys. Rev. Lett. \textbf{126}, 130501 (2021).

\bibitem{COMPACT4} L. He, D. Liu, J. Gao, W. Zhang, H. Zhang, X. Feng, Y. Huang, K. Cui, F. Liu, W. Zhang, and X. Zhang, {\it Super-Compact Universal Quantum Logic Gates with Inverse-Designed Elements}, Sci. Adva. \textbf{9}, eadg6685 (2023).



%%%%SWAP: Multiqubit swapping
\bibitem{SWAP1}	S. Bose, V. Vedral, and P. L. Knight, {\it Multiparticle Generalization of Entanglement Swapping}, Phys. Rev. A \textbf{57}, 822 (1998).

\bibitem{SWAP2}	C.-Y. Lu, T. Yang, and J.-W. Pan, {\it Experimental Multiparticle Entanglement Swapping for Quantum Networking}, Phys. Rev. Lett. \textbf{103}, 020501 (2009).
%%%%%%%%Book%%%%%%%%%%%%%%%%%%%%%%%%%
%\bibitem{BS-book}  E. Hecht, {\it Optics} (Pearson Education, Harlow, United Kingdom, 2013).

%\bibitem{QBS-book} G. S. Agarwal, {\it Quantum optics} (Cambridge University Press, New York, 2012).

\end{thebibliography}

\end{document}